# Kinetics of crystallization of FeB-based amorphous alloys studied by neutron thermo-diffractometry

A. Fernández-Martínez[a], P. Gorría[b,*], G. J. Cuello[a], J. D. Santos[b], M. J. Pérez[b]

[a]*Institut Laue-Langevin, B.P. 156, 38042 Grenoble, France*
[b]*Departamento de Física, Universidad de Oviedo, C/Calvo Sotelo s/n, 33007 Oviedo, Spain*

**Abstract**

Kinetics of crystallization of two amorphous alloys, $Fe_{70}Cr_{10}B_{20}$ and $Fe_{80}Zr_{10}B_{10}$, have been followed up by neutron thermodiffractometry experiments performed in the two axis diffractometer D20 (Institut Laue-Langevin, Grenoble). The structural changes are directly correlated with the temperature dependence of the magnetization. $Fe_{70}Cr_{10}B_{20}$ crystallizes following a two-step process: an eutectic crystallization of $\alpha$-Fe (bcc) and the metastable tetragonal phase $(Fe_{0.8}Cr_{0.2})_3B$ followed by another eutectic transformation to the stable phase $(Fe_{0.75}Cr_{0.25})_2B$ and more segregation of $\alpha$-Fe. These tetragonal phases are magnetically anisotropic, giving rise to a large increase of the coercivity. This behaviour is similar to that of $Fe_{80}B_{20}$ alloys, with Cr atoms replacing the Fe positions in both crystalline phases. $Fe_{80}Zr_{10}B_{10}$ shows also a two-step process in which two polymorphic transformations take place.

---

[*] Corresponding author: Tel: +34 985 102899 ; fax; +34 985 103324. E-mail address: pgorria@uniovi.es (P. Gorría).

Amorphous FeCrB and FeZrB alloys form part of a group of materials with both soft magnetic response at room temperature and anomalous magnetic behaviour like spin-glass, reentrant spin-glass and INVAR effect at low temperatures, especially in Fe-rich compounds. The addition of B to FeZr amorphous alloys increases the Curie temperature and the Fe magnetic moment [1]. Introduction of Cr changes the magnetic and mechanical properties of FeB-based metallic glasses, including higher resistance to corrosion [2-4]. Thermal treatments are usually carried out to get nanocrystalline or strain-relaxed states, which optimize the soft magnetic behaviour. They give rise to the appearance of nanocrystalline grains that overcome structural phase transitions when rising the temperature. These structural transitions are correlated to changes in the magnetic response of the materials. In this work, a neutron thermodiffraction experiment has been carried out in order to correlate the changes in the magnetization with structural phase transitions.

Ribbons of compositions $Fe_{70}Cr_{10}B_{20}$ and $Fe_{80}Zr_{10}B_{10}$ were synthesized by means of the melt spinning technique in a vacuum chamber. The magnetization vs. temperature curves, *M(T)*, were measured in a Faraday susceptometer under an applied magnetic field of 1 kOe and in the temperature range between 293 and 1073 K. In situ neutron thermodiffraction experiments were carried out at the two axis diffractometer D20 (ILL, Grenoble, France) in the same temperature range. The wavelength used was $\lambda \approx 1.3$ Å and the angular range, in $2\theta$, 0-157º. Differential scanning calorimetry (DSC) and hysteresis loops obtained from induction measurements [5] (not shown here) were performed to study both the crystallization process and the increase of coercivity with temperature due to magneto-crystalline anisotropy.

The thermomagnetic curves for both materials (Fig. 1) show that in the heating-up, the magnetization, *M*, of the as-quenched $Fe_{70}Cr_{10}B_{20}$ sample deeply decreases at around 390 K, indicating the Curie temperature, $T_C$, of the initial amorphous phase.

A sharp increase of *M* at above 723 K suggests the beginning of the crystallization of any ferromagnetic phase. Further heating leads to a progressive increase of *M* up to 1020 K, and then decreases. This indicates the presence of some amount of BCC-Fe ($T_C$ = 1043 K). On cooling down from high temperature, a change in the slope of M(T) curve is observed below 500 K, a signature of the Curie temperature of any ferromagnetic phase different from BCC-Fe. For the $Fe_{80}Zr_{10}B_{10}$ sample a similar behaviour is observed, with a sudden increase of *M* at about 975 K, a change in the slope at around 1000 K and a Curie temperature close to that of the BCC-Fe. On cooling down, the existence of a Curie temperature around 400 K is also evident.

Neutron thermodiffraction experiments for the $Fe_{70}Cr_{10}B_{20}$ sample show that the crystallization process is qualitatively identical to that of the $Fe_{80}B_{20}$ amorphous alloys [6]: an eutectic crystallization of BCC-Fe ($\approx 20\%$) and a tetragonal metastable phase with $I\bar{4}$ symmetry ($\approx 80\%$) takes place in a first stage at a temperature of 750 K. A composition for unit cell of $Fe_{20}Cr_4B_8$ for the tetragonal phase fits well in the Rietveld analyses of the diffraction patterns (see Fig. 2.) When the temperature rises up to around 920 K, another eutectic crystallization of BCC-Fe (up to 40%, thus explaining the increase of magnetization, see Fig. 1) and a tetragonal phase with *I4/mcm* symmetry and composition $Fe_6Cr_2B_4$ ($\approx$ 60%) takes place. BCC-Fe is segregated also from the first tetragonal phase. Hence, the two changes in the slope of the M(T) curve on cooling can now be identified with the $T_C$ for BCC-Fe and $(FeCr)_2B$ phases.

These two tetragonal phases are the same appearing during the eutectic crystallization of $Fe_{80}B_{20}$ metallic glasses, thus confirming the tendency of Cr to occupy the Fe atomic positions in FeCrB alloy.

In the case of $Fe_{80}Zr_{10}B_{10}$ the crystallization takes place following several primary transformations (see Fig. 3). The crystallization of FeZr mechanically-alloyed amorphous alloys has been already studied by different authors [7-10] and it seems to be very composition-dependent [7]. For the $Fe_{90}Zr_{10}$ alloy, a sequence of amorphous + BCC-Fe → BCC-Fe + FCC-$Fe_3Zr$ → BCC-Fe + FCC-$Fe_3Zr$ + FCC-$Fe_2Zr$ has been found [10]. In our case, the introduction of B modifies the process. Above 850 K, crystallization of the $Fe_2Zr$ phase with HCP structure takes place. This phase is paramagnetic at this temperature, because it is not found any change in *M* around 850 K. Heating above 975 K results in the segregation of BCC-Fe together with the beginning of the transformation between HCP and FCC crystal structures for the $Fe_2Zr$ phase. Also, some amount of metastable $Fe_3B$ phase with tetragonal crystal structure is formed. This phase has been found earlier in mechanically-induced phase transitions of $Fe_{70}Zr_{10}B_{20}$ [11]. Further heating above 1020 K leads to a supplementary segregation of BCC-Fe (reflected in an increase of *M*, see Fig. 1.) accompanied by the crystallization of another phase of FeB-type. Different candidates are hold for this phase, as its indexation is quite difficult since the peaks are broad and have tiny intensities. The more suitable phase would be an orthorhombic phase of equiatomic FeB. A similar phase has been found from Mössbauer spectroscopy during the crystallization of FeZrB-based metallic glasses [11,12]. Moreover, the M(T) curve, on cooling, suggest the existence of a ferromagnetic phase with a $T_C$ below 450 K (see Fig. 1), significantly lower to that of FeB (598 K [13]). However, some Zr atoms could be dissolved in such FeB phase, leading to a decrease of $T_C$. We can propose the following as the complete crystallization process for $Fe_{80}Zr_{10}B_{10}$: amorphous → amorphous + HCP-$Fe_2Zr$ → HCP-$Fe_2Zr$ + BCC-Fe + tetragonal-FeB + FCC-$Fe_2Zr$ → FCC-$Fe_2Zr$ + FeB + BCC-Fe.

In conclusion, these metallic glasses present complex crystallization processes. In the case of FeCrB alloy, the affinity of Cr for the Fe atomic positions in the crystallization products is shown. On the other hand, the crystallization behaviour for FeZrB is strong composition-dependent, exhibiting several primary transformations. A new crystallization sequence is proposed, with the appearance of the HCP-Fe$_2$Zr phase.

**Figure captions**

Fig. 1. Thermomagnetization curves for both alloys. The arrows show the crystallization temperatures of magnetic phases (see inset for $Fe_{80}Zr_{10}B_{10}$).

Fig. 2. Weight fraction diagram from the Rietveld refinement of the $Fe_{70}Cr_{10}B_{20}$ diffraction patterns. The crystallization of BCC-Fe at 750 K and 950 K explains the increase of *M* in Fig. 1.

Fig. 3. Neutron diffraction patterns showing the structural evolution of the $Fe_{80}Zr_{10}B_{10}$ sample at some selected temperatures.

Fig. 1.

A.Fernández-Martínez et al.

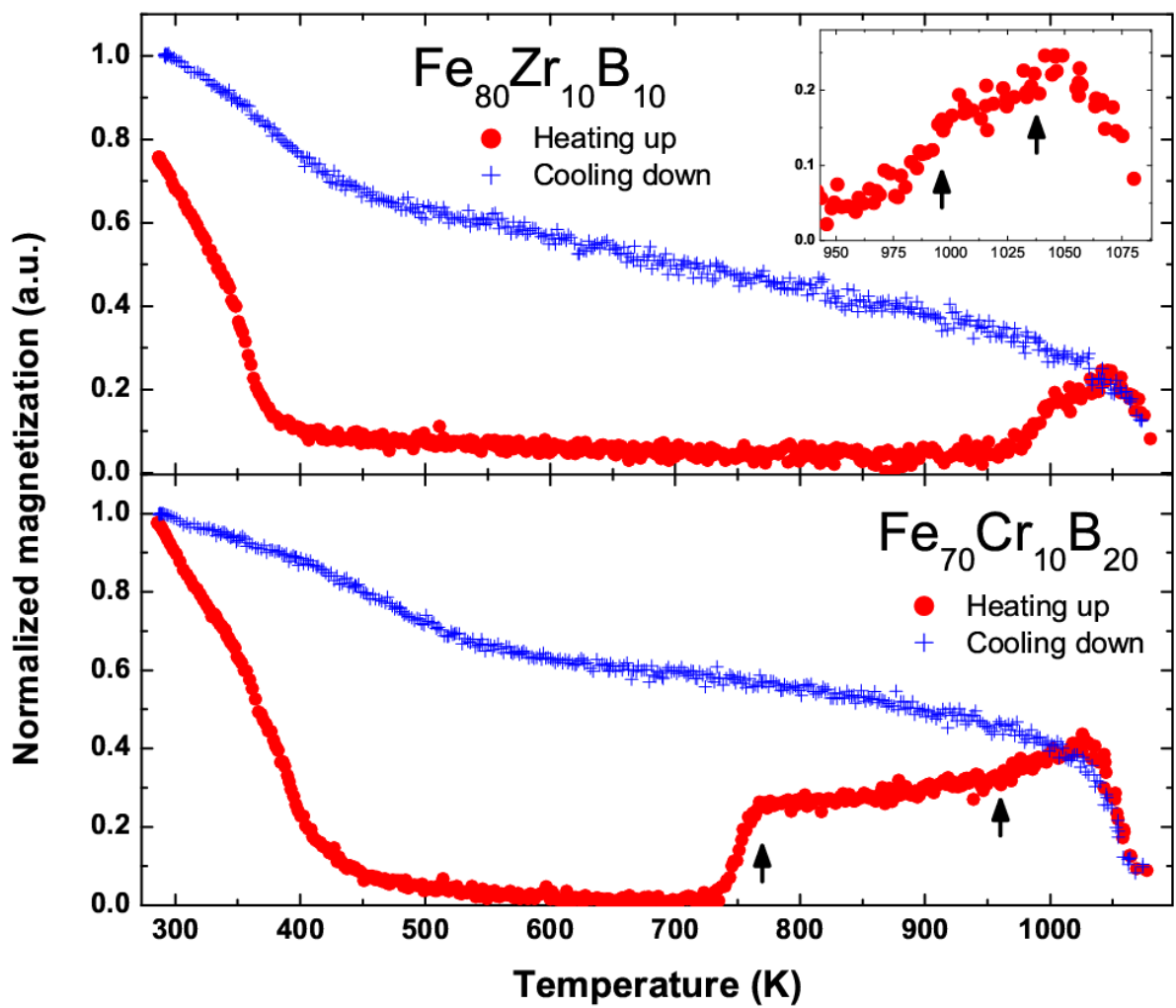

Fig. 2.

A.Fernández-Martínez et al.

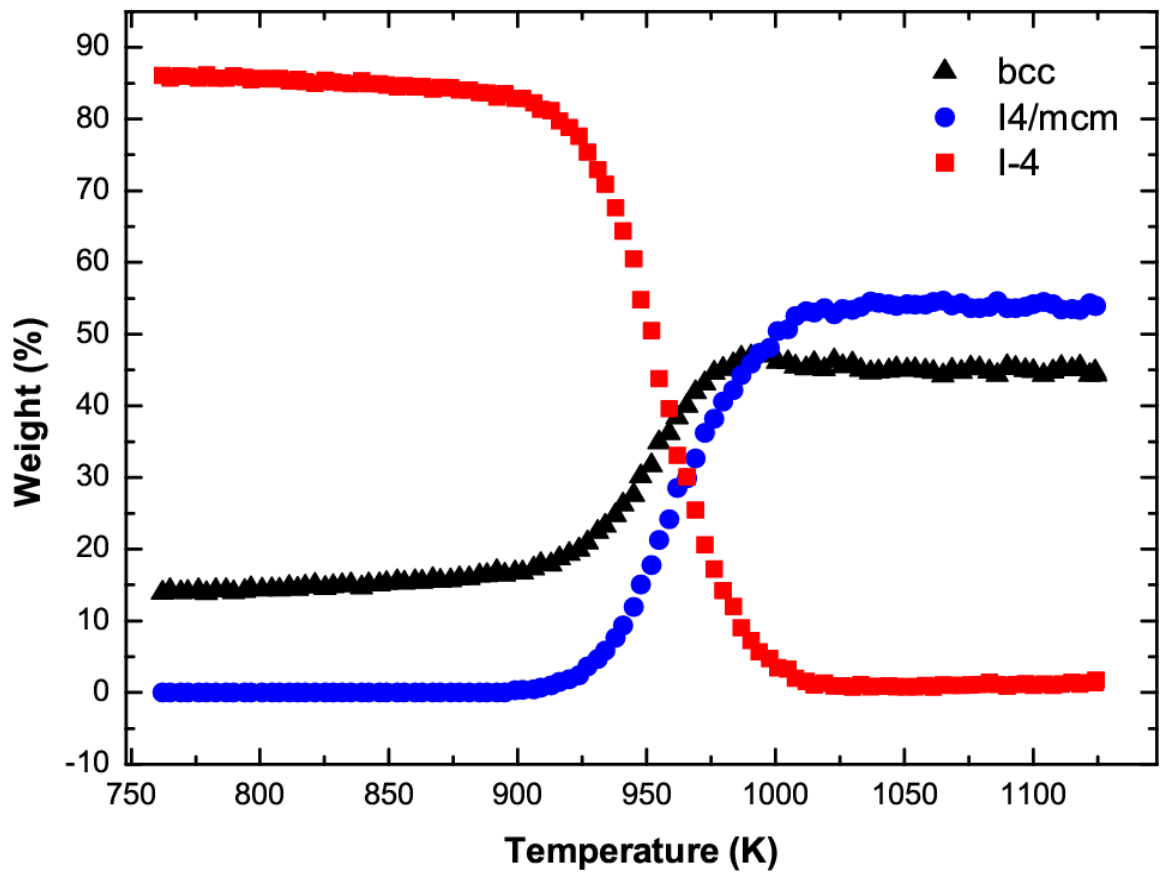

Fig. 3.

A.Fernández-Martínez et al.

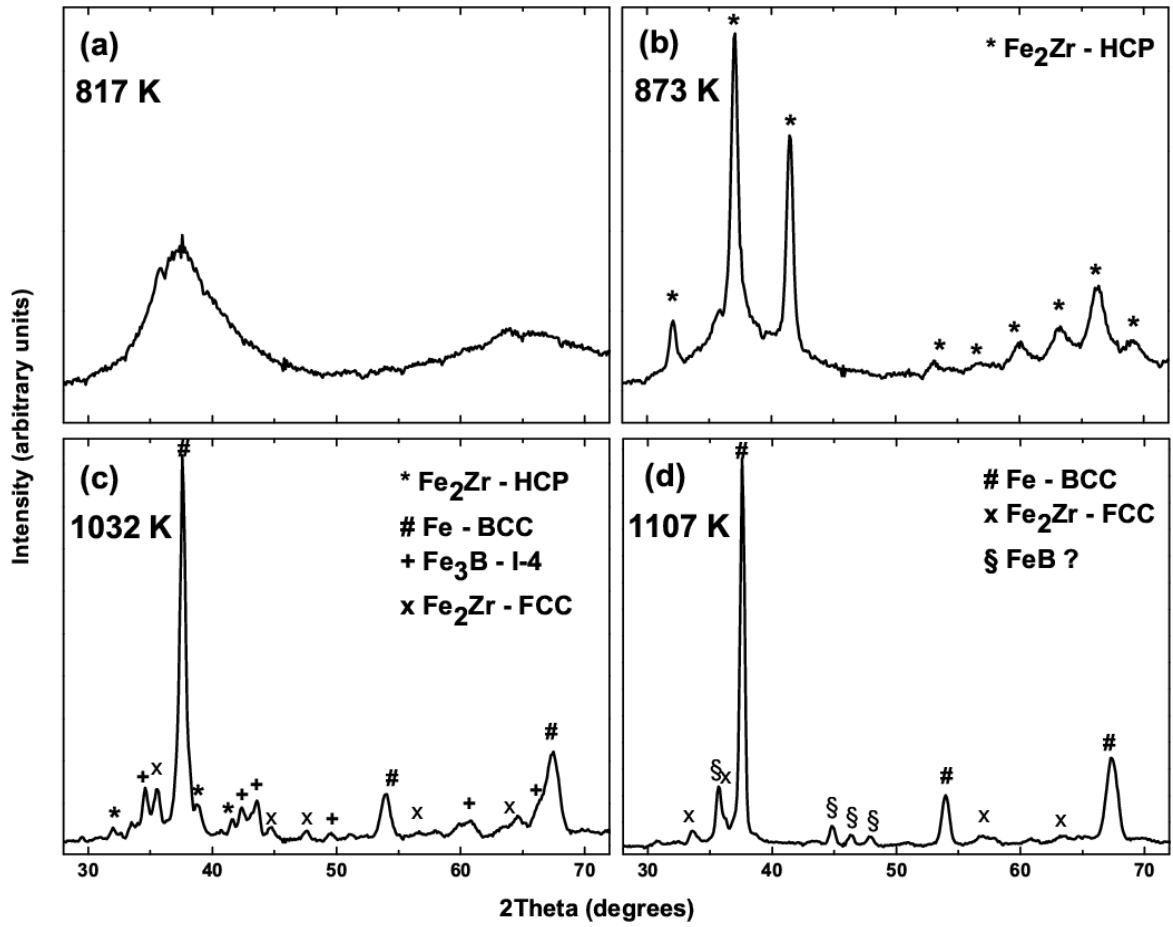